\pdfoutput=1
\documentclass[aps,pre,amsmath,amssymb,reprint]{revtex4-2}
\usepackage[utf8]{inputenc}
\usepackage[T1]{fontenc}
\usepackage{lmodern}
\usepackage{graphicx}
\usepackage{textcomp}
\usepackage[colorlinks=true,allcolors=blue]{hyperref}

\begin{document}

\title{Relative periodic orbits of a free spinning loop}
\author{Nicholas B. Tufillaro}
\email{nick@aqualytics.eco}
\affiliation{Aqualytics, Corvallis, Oregon, USA}

\begin{abstract}
A closed elastic string with no external forces can hold any shape while its material moves steadily along the curve (a steady axial motion), because its tension and the force needed to turn the material around a bend are both proportional to the curvature. I compute the periodic motions that grow from the simplest of these motions, the spinning ring, by Newton's method on a chain of masses. They include rotating waves that turn at twice the rotation rate of the material, standing ``flappers'' with two to five lobes, and period-doubled and period-tripled orbits. The chain only approximates the string. The orbits and their strong instabilities change little with more masses and are properties of the continuous string. Most weak instabilities change with the number of masses and are artifacts of approximating the string by a finite chain. The saddle flapper of a loop stretched 30\% beyond its natural length is neutrally stable at small heights and unstable at large ones. Near a stretch of 1/12 the saddle flapper resonates with the breathing of the ring. Below that stretch the flapper slows as it grows and above it the flapper speeds up, and near the resonance two families of saddle flappers overlap. An unstable saddle flapper trades energy back and forth with an in-plane oval, and at large amplitude the energy spreads to the shortest waves of the chain. For Keller's linear string, whose tension is proportional to the stretched length, every motion is periodic, and every knot carries an exactly periodic flapper. For a Hookean string the curvature of a knot couples a ripple to stretching waves that travel at a different speed, and in the simulations a ripple on a sliding knot does not return to its starting shape. Whether a Hookean knot carries any periodic flapper is open.
\end{abstract}

\maketitle

\section{Introduction}

A string whose material moves along its own curve at the speed of its transverse waves holds any shape. In 1878 Aitken demonstrated this ``rigidity produced by centrifugal force'' with chains spun on pulleys \cite{ref1,ref2}, and Routh showed that the tension of the moving chain supplies the centripetal force at every point of the curve, whatever its shape \cite{ref3}. Healey and Papadopoulos proved that a nonlinearly elastic string can move in this way along any closed curve \cite{ref4}. Healey proved that the simplest of these motions, the spinning circle, is stable when $T/\nu$ increases with the stretch, and found rigidly rotating noncircular loops that branch from the circle when it does not \cite{refH90}. Dehadrai and Hanna have recently reviewed this history and computed the shapes of driven loops under gravity and drag \cite{ref5}. As they note, one of the two arbitrarily shaped wave solutions of an axially moving string is frozen in the laboratory frame. In the engineering literature on axially moving continua the same condition appears as the critical speed, at which a backward transverse wave stands still in the laboratory \cite{ref6}. Exact large-amplitude motions of free strings, by contrast, are known only in special cases: every motion of Keller's linear string \cite{ref7}, the rotating and helical motions found by Rosenau and Rubin \cite{ref8,ref9}, and the geometric treatment of the linear string by Clelland and Vassiliou \cite{ref10}.

The sliding shapes are relative equilibria of a Hamiltonian system with symmetry, motions that are steady up to the action of a continuous group \cite{ref11}. Near a relative equilibrium one expects relative periodic orbits, which return to their starting state up to a fixed group element. For an equilibrium at which the energy has a definite second variation, the theorem of Weinstein and Moser gives periodic orbits on every nearby energy level \cite{refWein,refMos}, and Montaldi, Roberts and Stewart showed which periodic orbits branch from the linear modes of a symmetric Hamiltonian equilibrium \cite{ref12}. Near a relative equilibrium the motion splits into a drift along the group and a motion in a slice across it \cite{refRWL}, and Ortega proved that, under generic conditions, relative periodic orbits surround a relative equilibrium whose augmented energy has a definite second variation \cite{refOrt}. Wulff and Schilder proved that a smooth family of relative periodic orbits grows from a linear mode of a nondegenerate relative equilibrium when no other mode with its symmetry has the same frequency or a whole multiple of it \cite{refWS14}. These theorems are proved for a finite number of degrees of freedom. Van der Meer surveys the reduction of such systems near relative equilibria \cite{ref13}, and Golubitsky and Stewart classify the spatiotemporal symmetries of the orbits \cite{ref14}. Relative periodic orbits are now routinely computed by Newton's method in fluid turbulence \cite{ref15,ref16}. For Hamiltonian systems, Wulff and Schebesch continued relative periodic orbits by Gauss--Newton shooting, with the period and the drift among the unknowns \cite{refWS08}, and Wulff and Schilder followed their bifurcations \cite{refWS09}. Periodic orbits that grow from linear modes are the nonlinear normal modes of structural dynamics \cite{ref17}. Because the computation replaces the string by a chain of masses, I also test which orbits and which stability results change when the number of masses is increased. For a taut cable, Hill et al. found that energy passes between in-plane and out-of-plane modes only when coupling terms lock their phases \cite{ref18}. As far as I am able to determine, however, the relative periodic orbits about a sliding loop of string have not been computed, and their stability is unknown.

In this paper I compute and classify the periodic motions that grow from the simplest sliding loop, the spinning ring, for a free Hookean string with no gravity, air or contact. I use a chain of point masses, solve for relative periodic orbits by Newton's method, and obtain their stability from exact Floquet multipliers of a symplectic map. The orbits fall into three groups: relative equilibria (sliding shapes, the spinning ring and rotating waves), relative periodic orbits (standing and lopsided ``flappers,'' in which the ring bends out of its plane into $m$ lobes that rise and fall as it spins, and orbits that mix lobe numbers), and motions that stay near these orbits without repeating. I also test whether ripples on a sliding knot return periodically, and show that the answer depends on the tension law.

This paper is organized as follows. Section II states the model, its symmetries, the sliding relative equilibria and the linear spectrum of the spinning ring. Section III describes the numerical method. Section IV reports the orbits found, type by type, their stability, and which results survive refinement of the chain. Section V treats knots and the linear string. Section VI summarizes the results and discusses the motion at large amplitude. A companion paper gives a pedagogical derivation of the equations of motion, of the centripetal balance and of the in-plane frequencies of the ring, with problems for students \cite{ref19}.

\section{Model and symmetries}

\subsection{Equations of motion}

Let $s \in [0, L)$ be the natural (unstretched) arc length of a closed string of natural length $L$, and $\mathbf{r}(s, t)$ the position of the material point $s$ at time $t$. The stretch $\nu = |\mathbf{r}_{s}|$ is the ratio of stretched to natural length, and a stretch of 30\% means $\nu = 1.3$, a string 30\% longer than its natural length. The unit tangent is $\hat{\mathbf{t}} = \mathbf{r}_{s}/\nu$, and $\mu$ is the mass per unit natural length. For a perfectly flexible string whose tension depends only on the local stretch, $T = T(\nu )$, the equation of motion is \cite{ref20,ref21,ref22}

\begin{equation}
\begin{gathered}
\mu \, \mathbf{r}_{tt} = \frac{\partial}{\partial s} \left( \frac{T(\nu)}{\nu} \, \mathbf{r}_s \right) , \\
\nu = |\mathbf{r}_s| .
\end{gathered} \label{eq:1}
\end{equation}

I use Hooke's law, $T = K(\nu - 1)$, with $K = EA$ the stiffness, and in Sec. V the one-parameter family

\begin{equation}
T(\nu) = K(\nu - \lambda) , \label{eq:2}
\end{equation}

which is Hooke's law at $\lambda = 1$ and Keller's linear string at $\lambda = 0$ \cite{ref7}. In terms of the stretched arc length $\ell$ and the curvature $\kappa$ of the curve at fixed time, Eq.~(\ref{eq:1}) becomes $\rho_{s} \mathbf{a} = (\partial T/\partial \ell ) \hat{\mathbf{t}} + T\kappa \hat{\mathbf{n}}$, with $\mathbf{a} = \mathbf{r}_{tt}$, $\rho_{s} = \mu /\nu$ the mass per unit stretched length and $\hat{\mathbf{n}}$ the principal normal. A derivation from the Lagrangian is given in the companion paper \cite{ref19}.

Equation~(\ref{eq:1}) is Hamiltonian with energy $H = \int [\tfrac{1}{2}\mu |\mathbf{r}_{t}|^{2} + W(\nu )] ds$, $W' = T$. Its symmetry group is generated by time shifts, the Euclidean group $E(3)$ and, for a uniform loop, the relabelings $s \to s + c$, which form the circle group $SO(2)_{s}$. The conserved quantities are $H$, the momentum $\mathbf{P} = \int \mu \mathbf{r}_{t} ds$, the angular momentum $\mathbf{L} = \int \mu \mathbf{r} \times \mathbf{r}_{t} ds$ and the pseudomomentum $J = \int \mu \mathbf{r}_{t} \cdot \mathbf{r}_{s} ds$ \cite{ref23}, which is Healey's circulation of a moving loop up to a constant factor \cite{ref29}.

For numbers I use a steel wire of natural length 1 m and radius 0.29 mm, with $\mu = 2.114 \times 10^{-3}$ kg/m and $K = 52.84$ kN. In units of $L$ and $L(\mu /K)^{1/2} = 0.200$ ms the motion depends only on the stretch and the shape, and the results apply to any Hookean loop.

\subsection{Sliding relative equilibria}

Let $\mathbf{C}$ be any smooth closed curve parametrized so that $|d\mathbf{C}/ds| = \nu$ is constant. The motion $\mathbf{r}(s, t) = \mathbf{C}(s + ut/\nu )$ has acceleration $u^{2}\kappa \hat{\mathbf{n}}$, and Eq.~(\ref{eq:1}) is satisfied at every point if

\begin{equation}
T = \rho_s u^2 = \frac{\mu u^2}{\nu} . \label{eq:3}
\end{equation}

The curvature cancels from the normal balance, and every closed curve, knotted or not, slides steadily \cite{ref3,ref4,ref5}. Healey and Papadopoulos, who call these motions steady axial motions, proved this for closed space curves, including a loosely tied knot, and for nonlinearly elastic strings, and showed that among sliding loops of given angular momentum the circle spinning about its axis has the least energy \cite{ref4}. For Hooke's law the sliding speed is $u = [K\nu (\nu - 1)/\mu ]^{1/2}$, 3122 m/s at 30\% stretch. The motion is steady up to a relabeling and is a relative equilibrium of the $SO(2)_{s}$ action. A sliding loop has $\mathbf{P} = 0$, $\mathbf{L} = 2\rho_{s} u \mathbf{A}$ with $\mathbf{A}$ the vector area of $\mathbf{C}$, and $J = \mu u\nu L$. For the circle, relabeling coincides with rotation, and the sliding ring is a spinning ring with rotation rate $\omega = u/R$. At 30\% stretch $R = 0.207$ m and $2\pi /\omega = 0.416$ ms.

\subsection{Linear spectrum of the spinning ring}

On a straight string at stretch $\nu$, transverse and longitudinal waves travel relative to the material at

\begin{equation}
\begin{gathered}
c_t = \left( \frac{T}{\rho_s} \right)^{1/2} , \\
c_L = \nu \left( \frac{T'(\nu)}{\mu} \right)^{1/2} ,
\end{gathered} \label{eq:4}
\end{equation}

and for Hooke's law $c_{t}/c_{L} = [(\nu - 1)/\nu ]^{1/2}$. Equation~(\ref{eq:3}) states that $c_{t} = u$. Transverse waves on a sliding loop therefore travel at 0 and $2u$ in the laboratory, and every sliding loop has an infinite-dimensional family of neutral directions, the neighboring sliding shapes. For the ring the out-of-plane displacement $z(\phi , t)$, with $\phi$ the material angle, does not couple to stretching at linear order and obeys $z_{tt} = \omega^{2} z_{\phi \phi}$. An $m$-lobe bend $z \propto \cos (m\phi \mp m\omega t)$ has the laboratory frequencies

\begin{equation}
\Omega_{\rm out} = 0 \quad \text{and} \quad \Omega_{\rm out} = 2 m \omega , \label{eq:5}
\end{equation}

at every stretch. I call the bend $z = a \cos m\phi$ with the velocities of the ring the ``linear start.'' It is half a frozen pattern and half a pattern turning at $2\omega$, and in linear theory it returns $2m$ times per rotation.

In-plane bends couple to stretching, and their frequencies depend on the stretch. Linearizing Eq.~(\ref{eq:1}) about the ring in the frame that turns with the material shows that an in-plane wave with $m$ lobes, proportional to $\exp i(m\theta - \Omega_{\rm in} t)$ in the laboratory, has either zero frequency, a neighboring sliding shape, or a frequency that is a root of the cubic \cite{ref19}

\begin{equation}
\begin{aligned}
\Omega_{\rm in}^3 &- 4 m \omega \, \Omega_{\rm in}^2 + \left[ (5 - \beta) m^2 - \beta - 3 \right] \omega^2 \, \Omega_{\rm in} \\
 &+ 2 (\beta - 1) \, m (m^2 - 1) \, \omega^3 = 0 ,
\end{aligned} \label{eq:inplane}
\end{equation}

where $\beta = (c_{L}/c_{t})^{2} = \nu T'(\nu )/T$, which is $\nu /(\nu - 1)$ for Hooke's law. A negative root is a wave that turns against the rotation. For $m = 0$ Eq.~(\ref{eq:inplane}) gives the breathing of the ring at $\Omega_{\rm in} = (\beta + 3)^{1/2} \omega$, and for $m = 1$ its roots are 0 and $[2 \pm (2 + 2\beta )^{1/2}]\omega$. For $m \ge 1$ and $\beta > 1$ the left side is positive or zero at $\Omega_{\rm in} = 0$ and equals $-4(\beta + 1) m \omega^{3}$ at $\Omega_{\rm in} = 2m\omega$. Since it is negative for large negative $\Omega_{\rm in}$ and positive for large positive $\Omega_{\rm in}$, all three roots are real. With Eq.~(\ref{eq:5}), no small motion of the spinning ring of a string with $\beta > 1$ grows exponentially. For $\beta = 1$, the linear string of Sec. V, the roots are 0 and $2(m \pm 1)\omega$.

Healey proved more \cite{refH90}. He called a string stiff when $T/\nu$ increases with $\nu$, which is the condition $\beta > 1$, and showed by the energy-momentum method that the spinning ring of a stiff string is orbitally stable under the full equations of motion. For softer strings, rigidly rotating noncircular loops branch from the ring where $\beta (m^{2} - 3) = 1$, which is Eq.~(\ref{eq:inplane}) with $\Omega_{\rm in} = m\omega$, a shape at rest in the frame of the material. A Hookean string is stiff at every stretch. The condition $\beta > 1$ states that longitudinal waves are faster than transverse ones, and Carter and Martin found that a longitudinal wave speed no smaller than the transverse one is sufficient for the stability of rotating circular loops of cosmic string \cite{ref32}.

\begin{figure}
\includegraphics[width=\columnwidth]{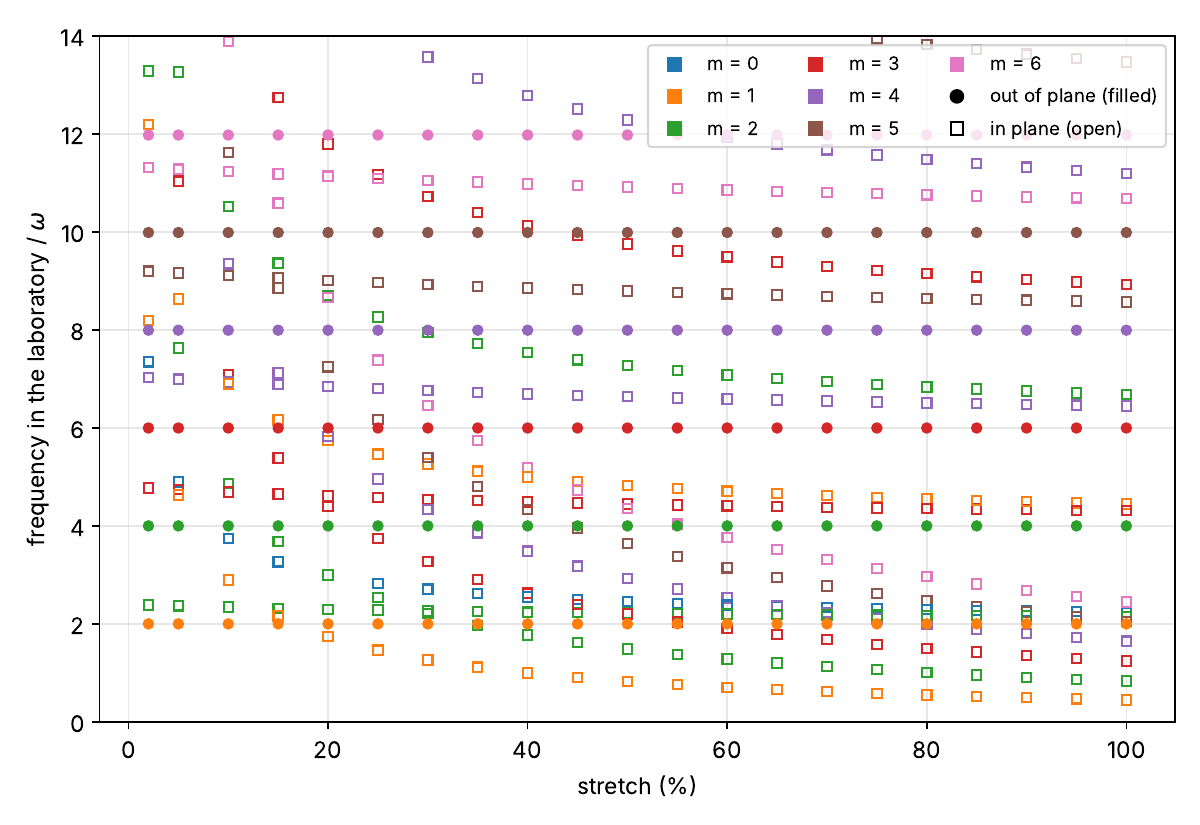}
\caption{Laboratory-frame frequencies of the spinning ring against stretch, in units of $\omega$ ($N = 120$). Filled circles, out of plane; open squares, in plane; color, lobe number $m$. Zero-frequency branches are omitted, and frequencies above $14\omega$ are not shown.}
\label{fig:1}
\end{figure}

Table~\ref{tab:I} and Fig.~\ref{fig:1} give the spectrum of the spinning ring computed from the exact Floquet matrix of the chain of Sec. III ($N = 120$) for stretches from 2\% to 100\%. The out-of-plane frequencies agree with Eq.~(\ref{eq:5}) to the accuracy of the chain ($3.999\omega$ for $m = 2$ and $7.993\omega$ for $m = 4$ at 30\% stretch). Each in-plane $m \ge 2$ has three nonzero branches, which fall with stretch and cross the out-of-plane lines. For $m$ up to 4 they agree with the magnitudes of the roots of Eq.~(\ref{eq:inplane}) to within 0.8\%. All multipliers lie on the unit circle to within $3 \times 10^{-6}$ per rotation, as the stability of the ring requires. The near-zero branches ($0.0008\omega$ to $0.024\omega$ for $m = 2$ to 6) are the chain's splitting of the continuum's degenerate sliding family and are omitted.

\begin{table*}[!htb]
\caption{Laboratory-frame frequencies of the small motions of the spinning ring, in units of $\omega$, for a chain of 120 masses. Out-of-plane values are the same at all stretches.}
\label{tab:I}
\begin{ruledtabular}
\begin{tabular}{lllll}
$m$ & Out of plane & In plane, 5\% & In plane, 30\% & In plane, 100\% \\
\hline
0 (breathing) & none & 4.90 & 2.71 & 2.24 \\
1 & 2.00 (tilt) & 4.63, 8.63 & 1.26, 5.26 & 0.45, 4.45 \\
2 & 4.00 & 2.37, 7.63, 13.26 & 2.22, 2.27, 7.95 & 0.83, 2.15, 6.68 \\
3 & 6.00 & 4.74, 11.03, 18.29 & 3.27, 4.54, 10.73 & 1.24, 4.31, 8.93 \\
4 & 7.99 & 6.99, 14.52, 23.52 & 4.33, 6.76, 13.57 & 1.65, 6.45, 11.19 \\
\end{tabular}
\end{ruledtabular}
\end{table*}

\subsection{Relative periodic orbits and their classification}

A relative periodic orbit of the free loop returns after a time $\tau$ to its initial state acted on by a rotation $R_{z}(\alpha )$ about the angular-momentum axis and a relabeling by $\Delta s$,

\begin{equation}
\begin{gathered}
\mathbf{r}(s, \tau) = R_z(\alpha) \, \mathbf{r}(s + \Delta s, 0) , \\
\dot{\mathbf{r}}(s, \tau) = R_z(\alpha) \, \dot{\mathbf{r}}(s + \Delta s, 0) .
\end{gathered} \label{eq:6}
\end{equation}

When $\Delta s$ is a rational fraction of $L$, as for every orbit computed here, and $\alpha /2\pi$ is irrational, the orbit fills a two-torus in phase space. In the space reduced by the symmetries and fixed values of the conserved quantities it is a periodic orbit, and on a Poincaré section it is a fixed point \cite{ref11,ref13}. Stability is read from the Floquet multipliers of the return map with the group action undone. The multipliers come in quartets $\Lambda$, $1/\Lambda$, $\Lambda^*$, $1/\Lambda^*$ \cite{ref24}, and those fixed at 1 by symmetries and conserved quantities are set aside. An orbit is elliptic if every remaining multiplier lies on the unit circle, hyperbolic if a real pair has $\Lambda > 1$, flip if a real pair has $\Lambda < -1$, and complex unstable if a quartet has left the circle in a Krein collision. A multiplier at $\exp (2\pi ip/q)$ marks a possible branch of period $q$. Where finer labels are needed I use the spatiotemporal symmetry pair $\mathcal{K} \subseteq \mathcal{H}$ of the H/K theorem (calligraphic letters, to distinguish them from the stiffness and the energy) \cite{ref14,ref25}.

The theorems cited in Sec. I bear on the chain of Sec. III, which has a finite number of degrees of freedom. Healey proved that the augmented energy $H - \omega L_{z}$ of the spinning ring of a stiff string has a definite second variation \cite{refH90}. For the ring of 120 masses its Hessian, with translations and the rotation about the axis removed, is positive definite at every stretch computed, from 2\% to 300\% (code and data in Ref.~\cite{refZenodo}), and Ortega's hypothesis holds for the chain. The nonresonance condition of Wulff and Schilder fails for the string. In the frame of the material the out-of-plane mode with $m$ lobes has frequency $m\omega$, and the modes with 6, 10, \ldots{} lobes have the symmetry of the saddle mode and odd multiples of its frequency. On the chain of 120 masses these resonances are only slightly detuned ($5.976\omega$ for six lobes against three times $1.9993\omega$). Near a stretch of 1/12 the condition fails again, since the breathing of the ring, which has every symmetry of the saddle mode, then has twice its frequency (Sec. IV~C). These theorems give relative periodic orbits near the ring of the chain, but they say nothing about which of them are stable or how far each family reaches, and they do not apply directly to the string. Section IV~F tests which of the computed orbits persist when the chain is refined.

\section{Numerical method}

The string is replaced by a closed chain of $N$ point masses $\mu L/N$ joined by springs that obey Eq.~(\ref{eq:2}), with natural spacing $h = L/N$,

\begin{equation}
\begin{gathered}
\frac{\mu L}{N} \, \ddot{\mathbf{r}}_j = T_j \, \hat{\mathbf{t}}_j - T_{j-1} \, \hat{\mathbf{t}}_{j-1} , \\
T_j = K\left( \frac{|\mathbf{r}_{j+1} - \mathbf{r}_j|}{h} - \lambda \right) ,
\end{gathered} \label{eq:7}
\end{equation}

with $\hat{\mathbf{t}}_{j}$ the unit vector from mass $j$ to mass $j + 1$. The chain keeps $H$, $\mathbf{P}$ and $\mathbf{L}$ exactly. It keeps the relabeling symmetry only for shifts by whole masses. For the ring a one-mass shift equals a rotation by $2\pi /N$, and the spinning ring is an exact relative equilibrium of the chain. For other curves it is not (Sec. V). I use $N = 120$ unless stated; Table~\ref{tab:II} lists the parameters.

\begin{table*}[!htb]
\caption{Parameters. Tension, speed, radius and period are for the spinning ring.}
\label{tab:II}
\begin{ruledtabular}
\begin{tabular}{lll}
Quantity & Symbol & Value \\
\hline
Natural length & $L$ & 1 m \\
Mass per unit natural length & $\mu$ & $2.114 \times 10^{-3}$ kg/m \\
Stiffness & $K = EA$ & 52.84 kN \\
Unit of time & $L(\mu /K)^{1/2}$ & 0.200 ms \\
Number of masses & $N$ & 120 (60 to 480 in checks, to 960 for knots) \\
Tension at 5\%, 30\%, 100\% stretch & $T$ & 2.64, 15.85, 52.84 kN \\
Sliding speed at 5\%, 30\%, 100\% & $u$ & 1146, 3122, 7071 m/s \\
Ring radius at 5\%, 30\%, 100\% & $R$ & 0.167, 0.207, 0.318 m \\
Rotation period at 5\%, 30\%, 100\% & $2\pi /\omega$ & 0.917, 0.416, 0.283 ms \\
Verlet time step & $\Delta t$ & about 0.4 \textmu{}s \\
\end{tabular}
\end{ruledtabular}
\end{table*}

Two integrators are used. Long runs use the energy-quadratization scheme of Ducceschi and Bilbao \cite{ref26} in the form used to simulate a whirling wire \cite{refWW}, with the damping and drive switched off and the ends joined. It conserves a discrete energy to round-off with one linear solve per step. Orbit searches use velocity Verlet \cite{ref27,ref28} with a fixed number $n$ of steps of length $\tau /n$, so that the time-$\tau$ map $\Phi_{\tau}$ is symplectic and smooth in $\tau$ as well as in the $6N$-dimensional state $\mathbf{x}$.

A relative periodic orbit, Eq.~(\ref{eq:6}), with $\Delta s$ equal to $k$ masses is a zero of

\begin{equation}
\begin{aligned}
\mathbf{F}(\mathbf{x}, \tau, \alpha) &= \Phi_\tau(\mathbf{x}) - R_z(\alpha)\, S_k\, \mathbf{x} \\
 &= 0 ,
\end{aligned} \label{eq:8}
\end{equation}

where $S_{k}$ shifts the labels by $k$. I append either $H = H_{0}$ and $L_{z} = L_{0}$, or $L_{z} = L_{0}$ and a fixed amplitude of one Fourier mode with $H$ left free, and solve by Gauss--Newton with minimum-norm steps (the pseudo-inverse of the Jacobian), which handles the neutral directions from time shifts, rotations and translations without explicit phase conditions \cite{ref15,ref16}. This is the Gauss--Newton shooting that Wulff and Schebesch developed for relative periodic orbits of Hamiltonian systems \cite{refWS08}. The Jacobian is formed by finite differences, with all $6N + 2$ perturbed states integrated together as one array. The residual $|\mathbf{F}|$ is measured with positions in units of $R$ and velocities in units of $u$. From a guess within a few percent Newton's method converges in four to six iterations for three quarters of the orbits, and the rest take up to 15. The residual ends below $10^{-10}$ for 96\% of the orbits and below $5 \times 10^{-9}$ for the rest, except the 4-lobe flapper at $0.275R$ ($2 \times 10^{-7}$ in the family run and $6 \times 10^{-11}$ when recomputed, with the same multipliers). Families are continued in amplitude in steps of $0.025R$, each converged orbit, scaled, seeding the next.

Rotating waves are relative equilibria of the continuum. On the chain they are relative periodic orbits of the one-mass shift, $\Phi_{\tau 1}(\mathbf{x}) = R_{z}(\alpha ) S_{1} \mathbf{x}$, with $\alpha /\tau_{1}$ the turning rate. They are seeded from a Floquet eigenvector of the ring and continued with $L_{z}$ held at the ring's value and the $m$-lobe amplitude fixed. Because the chain carries a nearby family of nearly steady sliding shapes (Sec. II C), Newton's method can converge onto that family instead. I accept a rotating wave only if its turning rate matches the linear value, $2\omega$ for out-of-plane waves by Eq.~(\ref{eq:5}) and, for in-plane waves, the frequency of the parent branch in Table~\ref{tab:I} divided by $m$.

Floquet multipliers are computed from the exact tangent map of the Verlet scheme, integrating $6N$ tangent vectors along the orbit and applying $(R_{z} S_{k})^{-1}$. The resulting matrix is symplectic to $4 \times 10^{-12}$. A finite-difference monodromy matrix splits the multipliers at 1 by about $10^{-4}$ and gives spurious weak instability. I count a multiplier as off the unit circle if $|\ln |\Lambda || > 10^{-5}$ per return.

To switch onto a branch of period $q$ at a multiplier near $\exp (2\pi ip/q)$, I add a small multiple of its eigenvector to the orbit, solve Eq.~(\ref{eq:8}) for the $q$-fold return, and append one equation fixing the amplitude of the new mode at a small value ($0.002R$), with $H$ free. This keeps Newton's method from falling back onto the parent orbit.

The orbits were checked with the other integrator. Saddle flappers at heights $0.05R$ and $0.1R$ run for 40 rotations with the energy-quadratization scheme ($N = 128$) repeat their peak heights to $2 \times 10^{-4}$, against $3 \times 10^{-3}$ and $10^{-2}$ for the linear start. Period-2 orbits hold their mode amplitudes over 20 rotations. Section IV~F repeats the orbit searches on chains of 60 to 480 masses. The MATLAB code runs in MATLAB and GNU Octave and is available from Zenodo \cite{refZenodo}, and eight movies are provided as Supplemental Material \cite{refSM}.

\section{Orbits of the spinning ring}

Unless stated otherwise the stretch is 30\%, heights are given in units of $R$ and the out-of-plane $m$-lobe amplitude is $Z_{m}$, the $m$th Fourier coefficient of $z$ around the axis, so that a bend $z = a \cos m\phi$ has $|Z_{m}| = a/2$. The in-plane amplitude is $R_{m}$. I call an instability strong if it grows by more than 25\% per rotation. Section IV F shows that strong growth is nearly the same on finer chains, while weaker growth often is not.

\subsection{Types of orbit}

\begin{figure*}
\includegraphics[width=\textwidth]{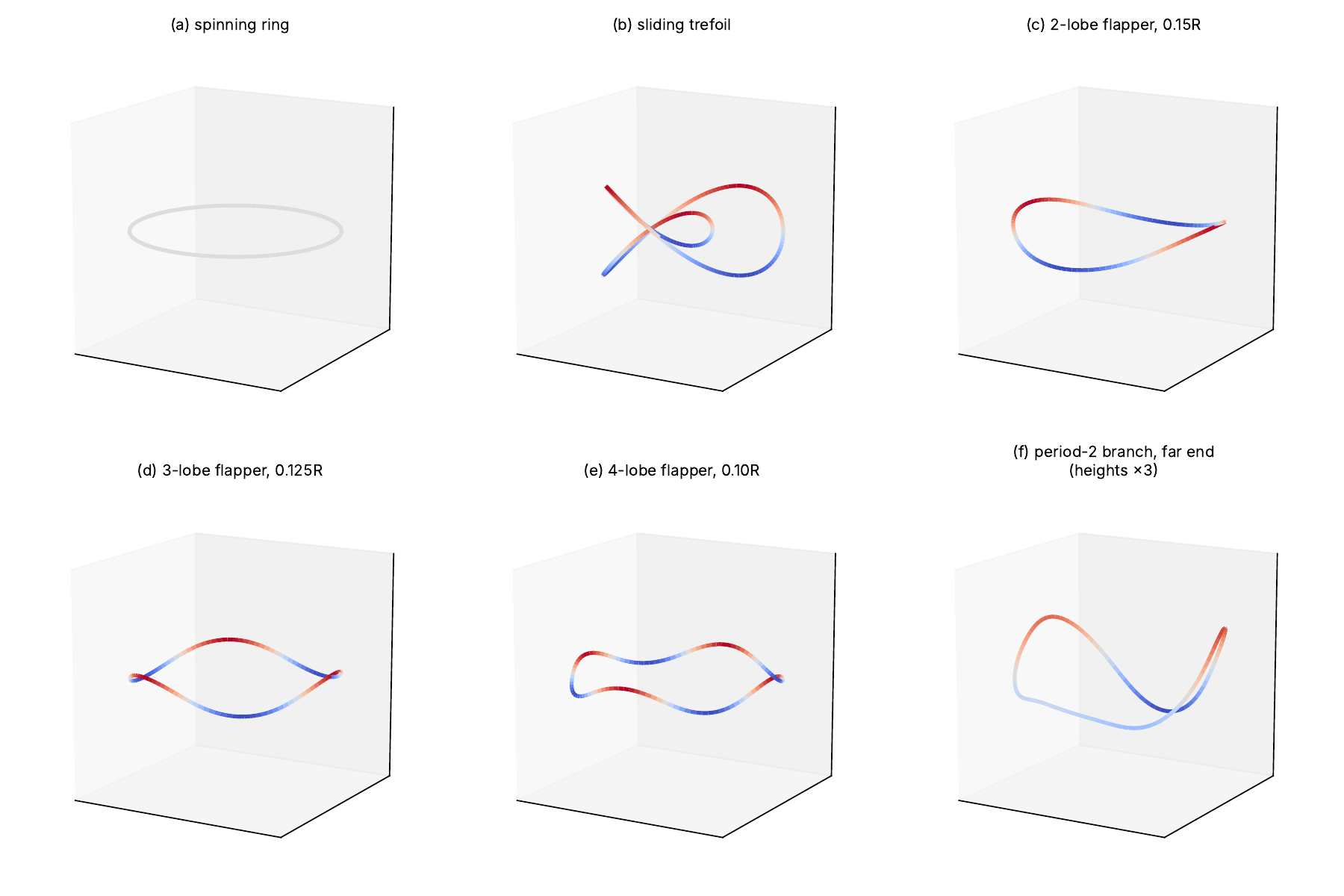}
\caption{Snapshots at 30\% stretch. (a) Spinning ring. (b) Sliding trefoil. (c)--(e) Standing flappers with $m = 2$, 3, 4 at maximum bend. (f) Far end of the period-2 branch, a nearly frozen saddle carrying a rotating 3-lobe wave, heights $\times 3$. Color, height above (red) or below (blue) the mean plane, scaled in each panel to its largest height.}
\label{fig:2}
\end{figure*}

Table~\ref{tab:III} lists the types found, ordered from relative equilibria to motion that stays near an orbit without repeating. For the out-of-plane families the type is fixed, at small amplitude, by how the frozen ($\Omega = 0$) and turning ($\Omega = 2m\omega$) waves of Eq.~(\ref{eq:5}) are mixed. A pure turning wave is a rotating wave. Equal parts give a standing flapper, whose lobes rise and fall in place; its $\mathcal{K}$ contains the rotations by $2\pi /m$ (kept at every instant) and its $\mathcal{H}$ the rotation by $\pi /m$, which exchanges hills and valleys and is kept after one flap. Unequal parts give a lopsided flapper, whose bend never passes through flat. Figure~\ref{fig:2} shows snapshots.

\begin{table*}[!htb]
\caption{Types of orbit of the free loop found in this work.}
\label{tab:III}
\begin{ruledtabular}
\begin{tabular}{p{0.293\textwidth}p{0.231\textwidth}p{0.291\textwidth}p{0.086\textwidth}}
\raggedright Type & \raggedright Defining condition & \raggedright Examples & \raggedright Sec. \tabularnewline
\hline
\raggedright Relative equilibrium: sliding shape & \raggedright Steady up to relabeling, Eq.~(\ref{eq:3}) & \raggedright Any closed curve; ring, trefoil, figure eight & \raggedright II B, V \tabularnewline
\raggedright Relative equilibrium: spinning ring & \raggedright Relabeling equals rotation & \raggedright Ring, all stretches; stable & \raggedright II C \tabularnewline
\raggedright Relative equilibrium: rotating wave & \raggedright Steady up to rotation at $\Omega \neq \omega$ and relabeling & \raggedright Out of plane, $m = 2$, 3, 4 at $\Omega \approx 2\omega$; in-plane oval at $\Omega \approx 4\omega$ & \raggedright IV B \tabularnewline
\raggedright Relative periodic orbit: standing flapper & \raggedright Eq.~(\ref{eq:6}), equal frozen and turning parts & \raggedright $m = 2$ to 5, stretches 5\% to 100\% & \raggedright IV C \tabularnewline
\raggedright Relative periodic orbit: lopsided flapper & \raggedright Eq.~(\ref{eq:6}), unequal parts & \raggedright Along the period-2 branch & \raggedright IV D \tabularnewline
\raggedright Relative periodic orbit: mixed branch & \raggedright Eq.~(\ref{eq:6}) after $q$ flaps & \raggedright Period 2 ($Z_{2}$ with $Z_{3}$); period 3 ($Z_{2}$ with $R_{3}$ or $R_{4}$) & \raggedright IV D \tabularnewline
\raggedright Motion near an orbit & \raggedright Stays close without repeating & \raggedright Linear start of the saddle; saddle--oval exchange; rippled knots & \raggedright IV C, IV E, V \tabularnewline
\end{tabular}
\end{ruledtabular}
\end{table*}

\subsection{Rotating waves}

The out-of-plane rotating waves with $m = 2$, 3 and 4 all turn at $2\omega$ to within 1.5\% over the amplitudes followed, as Eq.~(\ref{eq:5}) predicts for any $m$ (Table~\ref{tab:IV}), and the turning rate is the same with 120 and 240 masses ($2.000\omega$ at $0.005R$). At small amplitude the 3- and 4-lobe waves grow weakly, by 0.3\% to 1\% and 0.9\% to 2.5\% per rotation up to $|Z_{m}| = 0.008R$, and these rates agree between 120 and 240 masses (Sec. IV F). The weak growth of the 2-lobe wave, below 0.5\% per rotation up to $0.010R$, is not resolved, and above about $0.012R$ the computed waves of all three families change with the number of masses. Continuation fails at $|Z_{m}|$ of $0.022R$ to $0.026R$, where Newton's method moves onto the nearly sliding shapes of the chain. The in-plane oval that grows from the $7.95\omega$ branch of Table~\ref{tab:I} turns at $3.98\omega$ to $4.01\omega$ and is strongly unstable from the smallest amplitude computed. It grows by a factor of 1.6 per rotation at $0.005R$ and of 22 to 24 at $0.031R$, with both 120 and 240 masses. The logarithm of the growth per rotation is proportional to the amplitude, between $100|R_{2}|/R$ and $103|R_{2}|/R$ over this range, so the instability vanishes only as the amplitude goes to zero, where the ring is stable. Above $|R_{2}| = 0.031R$ its tension becomes negative in part of the loop, outside the model. Turning rates are in units of the ring's $\omega$, and growth is per rotation of the ring. At fixed $L_{z}$ the material of a wave turns slightly faster than the ring, by 1\% for the largest 4-lobe wave and 2\% for the largest oval. On the chain I also found a saddle-shaped sliding shape whose pattern turns at only $0.0003\omega$, a member of the nearly sliding family.

\begin{table*}[!htb]
\caption{Rotating waves at 30\% stretch, with $L_{z}$ equal to that of the ring. Turning rate in units of $\omega$; growth is the largest multiplier modulus per rotation of the ring, given only where it agrees between 120 and 240 masses.}
\label{tab:IV}
\begin{ruledtabular}
\begin{tabular}{lllll}
Family & Amplitude followed (R) & Turning rate ($\omega$) & Growth per rotation & Tension range (kN) \\
\hline
Out of plane, $m = 2$ & 0.005--0.022 & 2.000--2.005 & not resolved & 15.85--16.04 \\
Out of plane, $m = 3$ & 0.005--0.022 & 2.000--2.010 & 1.003--1.010 (to $0.008R$) & 15.86--16.23 \\
Out of plane, $m = 4$ & 0.005--0.026 & 1.999--2.027 & 1.009--1.025 (to $0.008R$) & 15.86--16.84 \\
In plane, $m = 2$ ($7.95\omega$ branch) & 0.005--0.031 & 3.98--4.01 & 1.6--23 & 1.5--31.1 \\
\end{tabular}
\end{ruledtabular}
\end{table*}

\subsection{Standing flappers}

\begin{figure*}
\includegraphics[width=\textwidth]{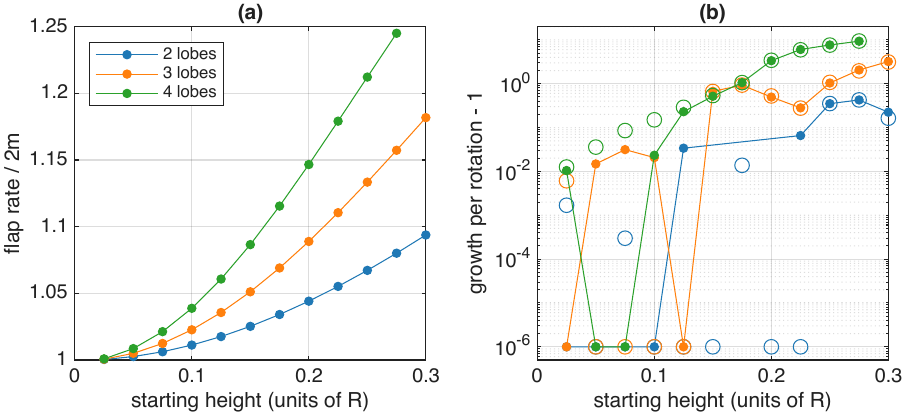}
\caption{Standing flappers at 30\% stretch, $m = 2$ to 4. (a) Flap rate (returns per rotation divided by $2m$) against starting height, 120 masses. (b) Largest multiplier modulus per rotation minus one, for 120 masses (open circles) and 240 masses (filled circles, joined); elliptic orbits are plotted at $10^{-6}$. Where open and filled circles disagree, the growth depends on the number of masses. No 240-mass points are shown for the 2-lobe flapper from $0.15R$ to $0.20R$, where the search stalled.}
\label{fig:3}
\end{figure*}

Each standing flapper family starts from the linear start, corrected by Newton's method to an orbit of Eq.~(\ref{eq:8}) with the same $H$ and $L_{z}$. At small amplitude a flapper returns $2m$ times per rotation, with $k = N/(2m)$ and $\alpha \approx 0$. With increasing height the flap rate rises (the flapper hardens) at most stretches. The 2-lobe flapper of height $0.3R$ returns 4.37 to 4.38 times per rotation (240 and 120 masses) at 30\% stretch (Fig.~\ref{fig:3}). From 10\% stretch up the shift decreases with stretch. For $m = 2$ at height $0.25R$ the flap rate exceeds $2m\omega$ by 19\% at 10\% stretch, 7\% at 30\%, 4\% at 50\% and 3\% at 100\%. At 5\% stretch the saddle softens, flapping 2.4\% below $2m\omega$ at $0.1R$. Hooke's law differs from the linear string, whose motions all share one period, only by the constant $K$ subtracted from the tension, and that constant matters less as the stretch grows. I conjecture that the decrease of the shift with stretch reflects the approach to that isochronous limit. In a two-mode model of a taut cable, Hill et al. find that modes not phase-locked to the response stiffen it by raising the axial tension \cite{ref18}, and a rise in the mean tension may likewise contribute to the hardening here.

The softening at 5\% stretch and the hardening from 10\% up lie on the two sides of a resonance. At second order the saddle drives the radius of the ring at its own return rate, $4\omega$, and by Eq.~(\ref{eq:inplane}) the ring breathes at $(\beta + 3)^{1/2}\omega$, which equals $4\omega$ at $\beta = 13$, a stretch of 1/12. I followed the saddle flapper in stretch at the energy and angular momentum of a linear start of height $0.05R$, and found two families near this stretch. In the first, continued up from 7\% stretch, $\alpha$ is positive and grows from 0.028 to 0.081 rad at 9.5\%, the flapper returns 3.94 to 3.81 times per rotation, and the ring is narrowest when the saddle is tallest. In the second, continued down from 10\% stretch, $\alpha$ is negative and falls from $-0.022$ to $-0.080$ rad at 7.3\%, the flapper returns 4.06 to 4.22 times per rotation, and the ring is widest when the saddle is tallest. This is the phase of an oscillator driven below and above its resonance. The two families overlap from 7.3\% to 9.5\% stretch. Along each family the saddle shrinks as the breathing grows, from 0.5\% to 1.8\% of $R$ (against 0.05\% at 30\% stretch), and each family ends on a breathing ring with no saddle, between 9.5\% and 9.6\% stretch for the first and between 7.2\% and 7.3\% for the second. On neither does $\alpha$ pass through zero. Every orbit computed on both families is elliptic. With 240 masses the two families are found again, also elliptic, with $\alpha$ within about 1\% (at most 1.6\%) and the returns per rotation within 0.001 of the 120-mass values (code and data in Ref.~\cite{refZenodo}).

\begin{figure*}
\includegraphics[width=\textwidth]{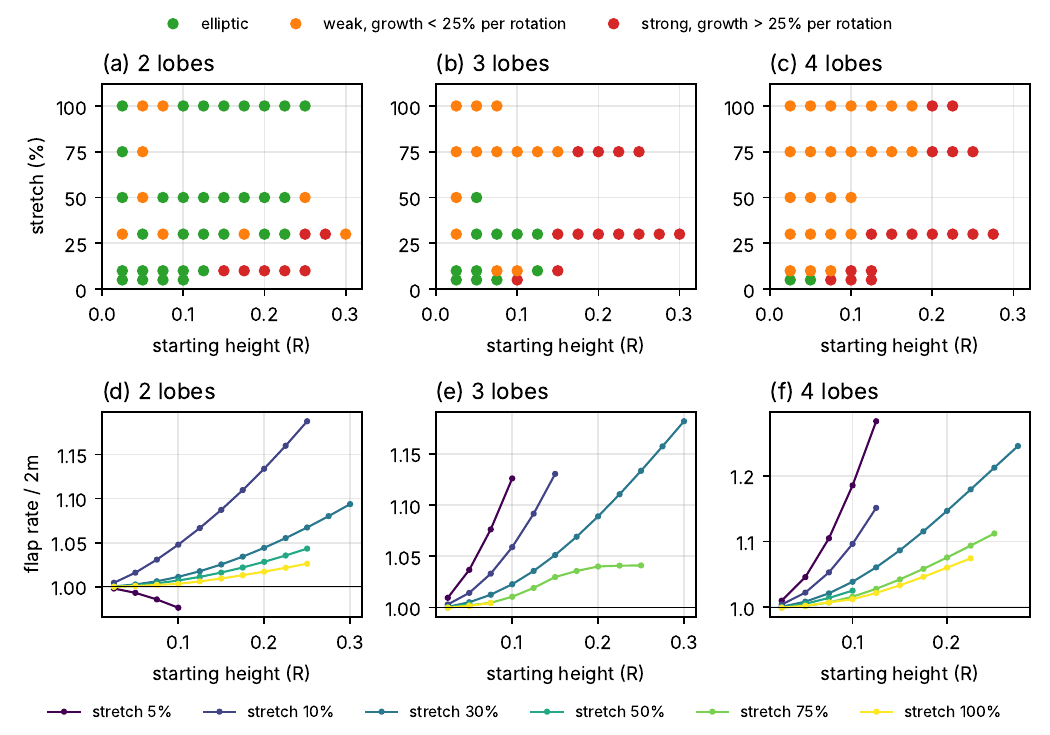}
\caption{Standing flappers against stretch. (a)--(c) Stability at each starting height (green, elliptic; orange, weak, growth below 25\% per rotation; red, strong), 120 masses; weak growth is mostly not resolved by the chain (Sec. IV F). (d)--(f) Flap rate divided by $2m\omega$ against starting height, one curve per stretch; horizontal line, linear value. Missing points, Newton failed.}
\label{fig:4}
\end{figure*}

I followed flappers with $m = 2$, 3, 4 at stretches 5\%, 10\%, 30\%, 50\%, 75\% and 100\%, and $m = 5$ at 30\%, with 120 masses (Fig.~\ref{fig:4}, Table~\ref{tab:V}). Table~\ref{tab:V} and Fig.~\ref{fig:4} sort the orbits into weak and strong instabilities by their largest growth per rotation, but weak growth is mostly not resolved by the chain: growth of up to 15\% per rotation with 120 masses falls below 3\% with 240 (Sec. IV F). At 30\% stretch the weak instabilities of the 2- and 3-lobe flappers, and at 100\% stretch those of the 2-lobe flapper, appear and disappear when the chain is refined. With that caveat, the 2-lobe flapper is neutrally stable at small and moderate heights at every stretch, and at 5\% stretch it is elliptic at every height computed with both 120 and 240 masses. At 10\% stretch it is strongly unstable from $0.15R$. At 30\% stretch it is strongly unstable at $0.25R$ and $0.275R$, with the same growth at $0.25R$ on chains of 60, 120 and 240 masses, and at $0.3R$ it grows by 16\% to 22\% per rotation, a weak instability that survives refinement (Sec. IV E). Flappers with three lobes are neutrally stable only at heights up to $0.125R$ and stretches up to 50\%, and flappers with four lobes only at 5\% stretch and heights up to $0.05R$. At 30\% stretch the 3- and 4-lobe flappers are strongly unstable from $0.15R$, again independent of the number of masses; at $0.125R$ the 4-lobe flapper grows by 29\% per rotation with 120 masses and by 23\% with 240. The 5-lobe flapper is unstable at every height computed, strongly so at $0.05R$ and from $0.125R$, with both 120 and 240 masses. Newton's method fails for some families at modest heights, from $0.075R$ for $m = 3$ at 50\% stretch and $m = 2$ at 75\%, from $0.1R$ for $m = 3$ at 100\%, and from $0.125R$ for $m = 2$ and 3 at 5\% and $m = 4$ at 50\%. The tension stays positive there (2.3 to 3.1 kN for $m = 2$ at 5\% and $0.1R$), and I suspect resonance with the in-plane branches of Fig.~\ref{fig:1}, which I have not checked.

\begin{table*}[!htb]
\caption{Stability of the standing flappers. Heights are starting heights in steps of $0.025R$. Growth is per rotation. All with 120 masses. Weak growth, below 25\% per rotation, mostly does not survive refinement of the chain (Sec. IV F). Last column, flap rate divided by $2m\omega$ at the largest height.}
\label{tab:V}
\begin{ruledtabular}
\begin{tabular}{p{0.094\textwidth}p{0.042\textwidth}p{0.148\textwidth}p{0.161\textwidth}p{0.148\textwidth}p{0.148\textwidth}p{0.094\textwidth}p{0.067\textwidth}}
\raggedright Stretch & \raggedright $m$ & \raggedright Heights found (R) & \raggedright Elliptic & \raggedright Weak, growth below 25\% & \raggedright Strong, growth above 25\% & \raggedright Largest growth & \raggedright Flap rate $/ 2m\omega$ \tabularnewline
\hline
\raggedright 5\% & \raggedright 2 & \raggedright 0.025--0.100 & \raggedright 0.025--0.100 & \raggedright none & \raggedright none & \raggedright 1.000 & \raggedright 0.976 \tabularnewline
\raggedright 5\% & \raggedright 3 & \raggedright 0.025--0.100 & \raggedright 0.025--0.075 & \raggedright none & \raggedright 0.100 & \raggedright 9.8 & \raggedright 1.126 \tabularnewline
\raggedright 5\% & \raggedright 4 & \raggedright 0.025--0.125 & \raggedright 0.025--0.050 & \raggedright none & \raggedright 0.075--0.125 & \raggedright 410 & \raggedright 1.283 \tabularnewline
\raggedright 10\% & \raggedright 2 & \raggedright 0.025--0.250 & \raggedright 0.025--0.125 & \raggedright none & \raggedright 0.150--0.250 & \raggedright 10.0 & \raggedright 1.188 \tabularnewline
\raggedright 10\% & \raggedright 3 & \raggedright 0.025--0.150 & \raggedright 0.025--0.050, 0.125 & \raggedright 0.075--0.100 & \raggedright 0.150 & \raggedright 5.0 & \raggedright 1.130 \tabularnewline
\raggedright 10\% & \raggedright 4 & \raggedright 0.025--0.125 & \raggedright none & \raggedright 0.025--0.075 & \raggedright 0.100--0.125 & \raggedright 19.5 & \raggedright 1.151 \tabularnewline
\raggedright 30\% & \raggedright 2 & \raggedright 0.025--0.300 & \raggedright 0.050, 0.100--0.150, 0.200--0.225 & \raggedright 0.025, 0.075, 0.175, 0.300 & \raggedright 0.250--0.275 & \raggedright 1.43 & \raggedright 1.094 \tabularnewline
\raggedright 30\% & \raggedright 3 & \raggedright 0.025--0.300 & \raggedright 0.050--0.125 & \raggedright 0.025 & \raggedright 0.150--0.300 & \raggedright 4.1 & \raggedright 1.182 \tabularnewline
\raggedright 30\% & \raggedright 4 & \raggedright 0.025--0.275 & \raggedright none & \raggedright 0.025--0.100 & \raggedright 0.125--0.275 & \raggedright 10.4 & \raggedright 1.245 \tabularnewline
\raggedright 30\% & \raggedright 5 & \raggedright 0.025--0.175 & \raggedright none & \raggedright 0.025, 0.075--0.100 & \raggedright 0.050, 0.125--0.175 & \raggedright 9.3 & \raggedright 1.169 \tabularnewline
\raggedright 50\% & \raggedright 2 & \raggedright 0.025--0.250 & \raggedright 0.025, 0.075--0.225 & \raggedright 0.050, 0.250 & \raggedright none & \raggedright 1.04 & \raggedright 1.043 \tabularnewline
\raggedright 50\% & \raggedright 3 & \raggedright 0.025--0.050 & \raggedright 0.050 & \raggedright 0.025 & \raggedright none & \raggedright 1.005 & \raggedright 1.003 \tabularnewline
\raggedright 50\% & \raggedright 4 & \raggedright 0.025--0.100 & \raggedright none & \raggedright 0.025--0.100 & \raggedright none & \raggedright 1.19 & \raggedright 1.025 \tabularnewline
\raggedright 75\% & \raggedright 2 & \raggedright 0.025--0.050 & \raggedright 0.025 & \raggedright 0.050 & \raggedright none & \raggedright 1.005 & \raggedright 1.001 \tabularnewline
\raggedright 75\% & \raggedright 3 & \raggedright 0.025--0.250 & \raggedright none & \raggedright 0.025--0.150 & \raggedright 0.175--0.250 & \raggedright 23.7 & \raggedright 1.041 \tabularnewline
\raggedright 75\% & \raggedright 4 & \raggedright 0.025--0.250 & \raggedright none & \raggedright 0.025--0.175 & \raggedright 0.200--0.250 & \raggedright 2.2 & \raggedright 1.112 \tabularnewline
\raggedright 100\% & \raggedright 2 & \raggedright 0.025--0.250 & \raggedright 0.025, 0.100--0.250 & \raggedright 0.050--0.075 & \raggedright none & \raggedright 1.007 & \raggedright 1.026 \tabularnewline
\raggedright 100\% & \raggedright 3 & \raggedright 0.025--0.075 & \raggedright none & \raggedright 0.025--0.075 & \raggedright none & \raggedright 1.03 & \raggedright 1.004 \tabularnewline
\raggedright 100\% & \raggedright 4 & \raggedright 0.025--0.225 & \raggedright none & \raggedright 0.025--0.175 & \raggedright 0.200--0.225 & \raggedright 1.65 & \raggedright 1.074 \tabularnewline
\end{tabular}
\end{ruledtabular}
\end{table*}

The linear start is not periodic. Started at $0.05R$ to $0.2R$ its peak heights wander by 0.3\% to 3\%, roughly as the square of the height, and it stays close to the flapper, as it would on a thin torus about it.

\subsection{Orbits branching from the saddle flapper}

I searched each 2-lobe flapper at 30\% and 100\% stretch for multipliers within 0.004 of $-1$ or $\exp (\pm 2\pi i/3)$ and attempted a branch switch at each. Of 17 candidates, three converged, all at 30\% stretch.

\begin{figure}
\includegraphics[width=\columnwidth]{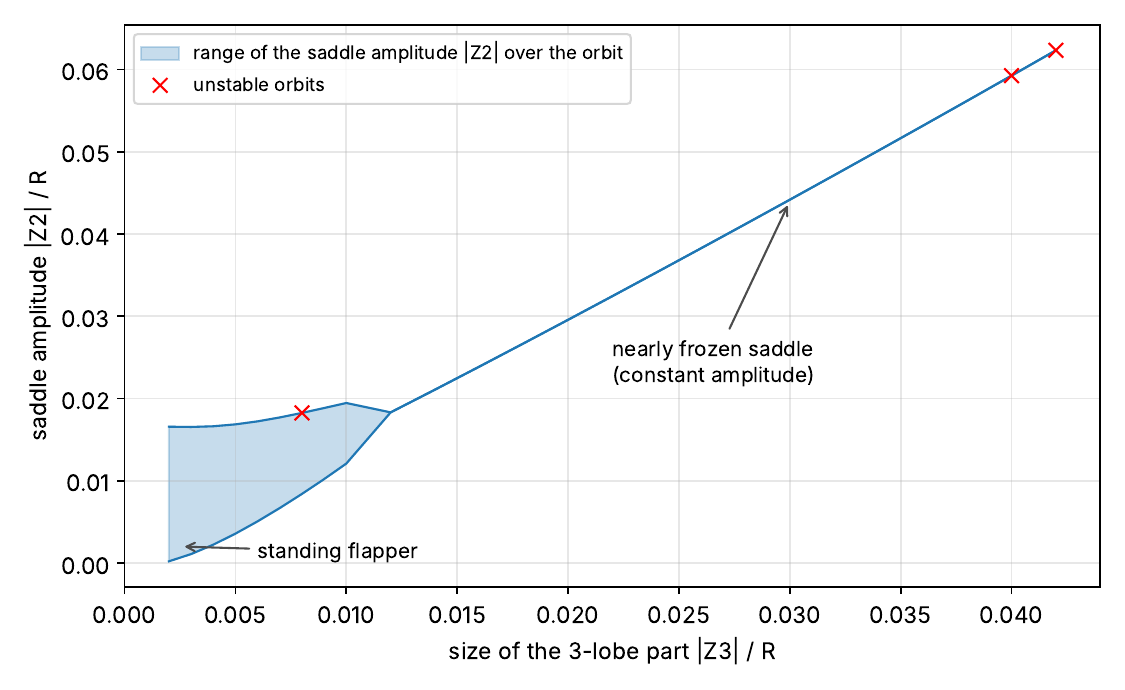}
\caption{Period-2 branch from the 2-lobe flapper at $0.05R$, 30\% stretch. Shaded band, range of $|Z_{2}|$ over one orbit against $|Z_{3}|$; crosses, unstable orbits. The band closes near $|Z_{3}| = 0.012R$, where the saddle stops flapping and stands nearly frozen in the laboratory.}
\label{fig:5}
\end{figure}

The first is a period-doubled orbit found from the flapper at $0.05R$, at a multiplier $\exp (2\pi i \times 0.4994)$ whose eigenvector is an out-of-plane 3-lobe mode. This multiplier is not special to that height. By Eq.~(\ref{eq:5}) a turning wave with $m'$ lobes advances by $2\pi m'/m$ per return of an $m$-lobe flapper, so that in linear theory every out-of-plane wave with an odd number of lobes has the multiplier $-1$ on the saddle flapper. The 3-lobe pair lies at $\exp (2\pi i \times 0.4998)$, $\exp (2\pi i \times 0.4994)$ and $\exp (2\pi i \times 0.4990)$ for the flappers at $0.025R$, $0.05R$ and $0.075R$. The period-doubled orbits are therefore the nonlinear continuation of a mixture of 2- and 3-lobe waves that exists, in linear theory, at any small height, and the branch followed here is one path through that family. It was continued in $|Z_{3}|$ from $0.002R$ to $0.042R$ (25 orbits, $|\mathbf{F}| < 10^{-10}$) at fixed $L_{z}$ with $H$ free, the energy rising from 12706 J to 14283 J. Its first orbit lies near a flapper of height about $0.034R$, with less energy than the flapper at $0.05R$ (12734 J). Along the branch the 3-lobe part keeps a constant size and turns through half a revolution per flap, which exchanges its hills and valleys, and the saddle stops flapping (Fig.~\ref{fig:5}). At $|Z_{3}| = 0.002R$ the saddle amplitude $|Z_{2}|$ swings between $0.0002R$ and $0.017R$, a standing flapper. At $0.010R$ it swings between $0.012R$ and $0.020R$, a lopsided flapper. From about $0.012R$ it is constant, and the saddle is then nearly frozen in the laboratory: its pattern drifts backward at $0.005\omega$ to $0.05\omega$, while the 3-lobe wave turns at $2.01\omega$ to $2.11\omega$. The branch therefore connects the standing flapper, through lopsided flappers, to a nearly frozen saddle, a sliding shape, that carries a rotating 3-lobe wave. The shape changes continually and repeats every two flaps. It is elliptic except for a weak bubble at $|Z_{3}| = 0.008R$ ($|\Lambda | = 1.00016$ per return) and from $0.040R$ ($|\Lambda | = 1.0007$ per return, rising to 1.009 at $0.042R$). Newton's method fails at $0.044R$. With 240 masses the branch is elliptic at $|Z_{3}| = 0.006R$, $0.02R$ and $0.038R$, and its returns per rotation change by about 0.001.

The other two are period triplings. From the flapper at $0.125R$ an in-plane 3-lobe mode gives an orbit in which the saddle flaps ($|Z_{2}|$ from $0.0015R$ to $0.064R$) with an in-plane 3-lobe ripple of $0.0020R$ to $0.0022R$, repeating every three flaps. From the flapper at $0.175R$ an in-plane 4-lobe mode gives a flapping saddle ($|Z_{2}|$ from $0.0038R$ to $0.088R$) with a flapping 4-lobe ripple ($0.0002R$ to $0.0028R$). With 120 masses both are weakly unstable, with $|\Lambda | = 1.018$ and 1.010 per three flaps. On finer chains the first stays weakly unstable but its growth doubles, from 2.4\% to 5.0\% per rotation with 240 masses, and the second is elliptic with 160 and 200 masses. The stability of both is therefore unresolved, and neither was continued. The remaining 14 candidates (five at 30\% stretch and all nine at 100\%) gave no converged orbit. In most of them the initial residual was $10^{2}$ to $10^{8}$, because the eigenvector's projection on the fixed amplitude was small and the push was scaled too large; normalizing the push by the eigenvector's own size should cure this.

\subsection{Saddle--oval exchange near an unstable flapper}

\begin{figure}
\includegraphics[width=\columnwidth]{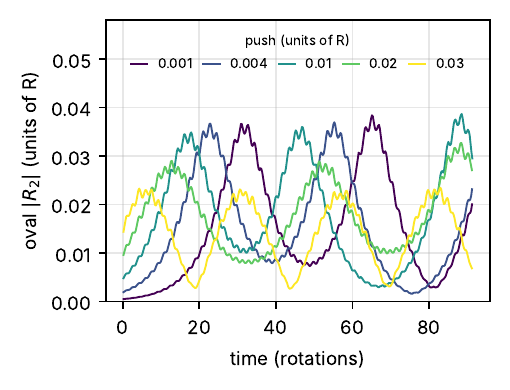}
\caption{Saddle--oval exchange. The 2-lobe flapper at $0.3R$ (30\% stretch) is pushed by $0.001R$ to $0.03R$ along its unstable eigenvector; each curve shows $|R_{2}|$ at successive maxima of the saddle against time in rotations.}
\label{fig:6}
\end{figure}

The 2-lobe flapper at $0.3R$ is complex unstable, with a leading multiplier of modulus 1.035 to 1.047 per flap (120 and 240 masses) and angle $2\pi \times 0.43$, whose eigenvector is an in-plane 2-lobe (oval) mode. Pushed along it by $0.001R$ to $0.03R$ and run for 300 to 400 flaps, the loop does not depart far. With 120 masses the oval amplitude $|R_{2}|$ grows to $0.035R$ to $0.038R$ while the saddle's peak height drops by a few percent, then falls back to $0.002R$ to $0.008R$, recurring every 130 to 150 flaps (30 to 35 rotations, Fig.~\ref{fig:6}). With 240 masses the same exchange occurs, with the oval reaching $0.050R$ and falling back to $0.002R$, but the growth rate changes by a third and the timing by a fifth. The chain therefore establishes that the exchange occurs, but not its rate or timing. The minima differ from cycle to cycle, and the motion does not repeat exactly. An exactly periodic exchange orbit would have a period of order 150 flaps and was not sought. The flapper at $0.275R$ behaves differently. Its multiplier, 1.086 per flap, is the same with 120 and 240 masses, and the oval it releases passes half its maximum within 14 rotations, grows to $0.09R$ to $0.10R$, and does not return within 90 rotations. For a taut cable, Hill et al. show that internal resonance between in-plane and out-of-plane modes requires phase-locking coupling terms \cite{ref18}. Whether the exchange here is the beating of such a near-resonant pair, close to a crossing of the in-plane and out-of-plane branches of Fig.~\ref{fig:1}, I have not determined.

\subsection{Convergence under refinement of the chain}

All of the orbits above were computed with 120 masses. To separate properties of the string from properties of the chain, I resampled each orbit by Fourier interpolation to chains of 60, 160, 200, 240 and (for one orbit) 480 masses, scaled the time step with the mass spacing, and solved Eq.~(\ref{eq:8}) again. Where a direct jump from 120 to 240 masses failed, I stepped through 160 and 200. Stepping did not always succeed. For the saddle flappers from $0.15R$ to $0.20R$ and the period-3 orbit with an in-plane 4-lobe ripple, Newton's method converged with 160 and 200 masses, where these orbits are elliptic, and stalled with 240 ($|\mathbf{F}|$ from $6 \times 10^{-4}$ to $2 \times 10^{-2}$). The 2-lobe rotating wave above $0.012R$ stalled with 240 masses or, at larger amplitude, with fewer. For the 3- and 4-lobe rotating waves Newton's method converged at each step, but the waves found with 240 masses drift away from those found with 120 as the amplitude grows: the 4-lobe wave at $0.026R$ turns at $2.12\omega$ with 240 masses, against $2.03\omega$ with 120. I do not use the multipliers of any of these orbits at 240 masses. Table~\ref{tab:VI} gives representative results.

The orbits themselves converge. The saddle flapper at $0.05R$ returns 4.0062, 4.0103, 4.0113 and 4.0116 times per rotation with 60, 120, 240 and 480 masses, and the differences fall by factors of 3.9 and 3.4 at each doubling, near the 4 expected of the $O(h^{2})$ error of the chain. The return rates of the other flappers agree to within 0.3\% between 120 and 240 masses at 30\% stretch, and to within 0.2\% at 100\% stretch. Strong instabilities converge as well: the largest multiplier of the saddle flapper at $0.25R$ is 1.0749, 1.0734 and 1.0730 per flap with 60, 120 and 240 masses.

Weak instabilities mostly do not converge. The weak complex instabilities of the 2-lobe flapper, at 30\% and at 100\% stretch, appear and disappear as the number of masses changes, and so does the growth of the period-3 orbit with an in-plane 4-lobe ripple. Such growth is not always small. The 4-lobe flapper at $0.1R$ grows by 15\% per rotation with 120 masses and by 2\% with 240, and the 2-lobe flapper at $0.225R$ is elliptic with 120 masses and grows by 7\% per rotation with 240. The weak growth of the 2-lobe rotating wave is not resolved. Each is most likely a collision between a mode of the orbit and a short wave of the chain, whose frequency moves when the chain is refined; I have not checked this mode by mode. The weak growth of the 3- and 4-lobe rotating waves at small amplitude is the exception, and it agrees closely between 120 and 240 masses. At $0.0124R$ it already falls as the chain is refined (for the 4-lobe wave 1.050, 1.047, 1.041 and 1.024 per rotation with 120, 160, 200 and 240 masses), and above that amplitude it is not resolved. Growth of 35\% per rotation or more agreed between 120 and 240 masses to within about 3\% for every flapper with two to four lobes, and to within 6\% for the 5-lobe flappers. Between 25\% and 35\% the agreement is poorer: 28\% with both chains for the 3-lobe flapper at $0.225R$, 33\% and 30\% for the 5-lobe flapper at $0.05R$, and 29\% and 23\% for the 4-lobe flapper at $0.125R$. I therefore treat growth of less than about 25\% per rotation as a property of the string only where it survives refinement.

\begin{table*}[!htb]
\caption{Convergence of representative orbits at 30\% stretch (5\% where stated). Return rate per rotation (turning rate in units of $\omega$ for rotating waves) and growth per rotation, with 120 and 240 masses.}
\label{tab:VI}
\begin{ruledtabular}
\begin{tabular}{lll}
Orbit & Return or turning rate, 120 / 240 & Growth per rotation, 120 / 240 \\
\hline
Saddle flapper, $0.05R$ & 4.0103 / 4.0113 & 1.000 / 1.000 \\
Saddle flapper, $0.125R$ & 4.0703 / 4.0716 & 1.000 / 1.034 \\
Saddle flapper, $0.25R$ & 4.2689 / 4.2702 & 1.353 / 1.351 \\
Saddle flapper, 5\% stretch, $0.10R$ & 3.9047 / 3.9055 & 1.000 / 1.000 \\
3-lobe flapper, $0.15R$ & 6.3070 / 6.3119 & 1.665 / 1.672 \\
4-lobe flapper, $0.20R$ & 9.1729 / 9.1880 & 4.43 / 4.35 \\
Period-2 orbit, $|Z_{3}| = 0.038R$ & 2.1313 / 2.1319 & 1.000 / 1.000 \\
Period-3 orbit, in-plane 3-lobe ripple & 1.3592 / 1.3620 & 1.024 / 1.050 \\
Rotating wave, out of plane, $m = 3$, $0.0072R$ & 2.0001 / 2.0009 & 1.008 / 1.007 \\
Rotating wave, in plane, $m = 2$, $0.005R$ & 3.976 / 3.978 & 1.649 / 1.646 \\
\end{tabular}
\end{ruledtabular}
\end{table*}

\section{Knots and the linear string}

By Eq.~(\ref{eq:3}) a trefoil or figure-eight knot slides as readily as a ring. I used the trefoil ($\sin \sigma + 2 \sin 2\sigma$, $\cos \sigma - 2 \cos 2\sigma$, $-\sin 3\sigma$) and the figure eight ($(2 + \cos 2\sigma ) \cos 3\sigma , (2 + \cos 2\sigma ) \sin 3\sigma , \sin 4\sigma$), scaled to the stretched length and sampled at equal arc length, with each mass given the sliding velocity $u\hat{\mathbf{t}}$. At 30\% stretch the trefoil's shape drifts by $0.0053R$ over three trips of the material with $N = 240$, and by $0.0016R$ over 1.1 trips with $N = 480$. Released from rest, the stretched trefoil collapses and passes through itself, since the model has no contact.

On the chain a sliding knot is not an exact solution. A one-mass shift is not a symmetry of the discrete knot, and the continuum's infinite-dimensional family of sliding shapes is broken by the discretization. A Newton search for a discrete sliding trefoil (a relative periodic orbit of the one-mass shift) does not converge: $|\mathbf{F}|$ stalls near $10^{-3}$ for $N = 60$, 120 and 240 while the shape wanders by $0.07R$ to $0.4R$. A Newton search for an exactly periodic rippled trefoil ($N = 120$) stalls at $|\mathbf{F}| = 0.16$ to 0.22.

I therefore tested whether ripples on a sliding knot return, by direct simulation. Each mass is displaced along the binormal by $0.03R \cos (12\pi s/L)$, six waves around the knot, keeping the sliding velocity of the unrippled knot. The return is measured by $D(t)$, the root-mean-square distance of the masses from the starting curve.

The reference case is Keller's linear string, $\lambda = 0$ in Eq.~(\ref{eq:2}) \cite{ref7,ref10}. Then $T/\nu = K$ and Eq.~(\ref{eq:1}) is the linear vector wave equation

\begin{equation}
\mu \, \mathbf{r}_{tt} = K \, \mathbf{r}_{ss} , \label{eq:9}
\end{equation}

with solutions $\mathbf{r} = \mathbf{f}(s + ct) + \mathbf{g}(s - ct)$, $c = (K/\mu )^{1/2}$. Every motion of a closed linear string with its center of mass at rest has period $L/c$, and a ripple $\mathbf{d}$ on a sliding curve $\mathbf{C}$, $\mathbf{r} = \mathbf{C}(s + ct) + \mathbf{d}(s + ct)/2 + \mathbf{d}(s - ct)/2$, returns exactly after half a trip. Every closed curve of the linear string, knotted or not, carries a periodic flapper.

\begin{figure*}
\includegraphics[width=\textwidth]{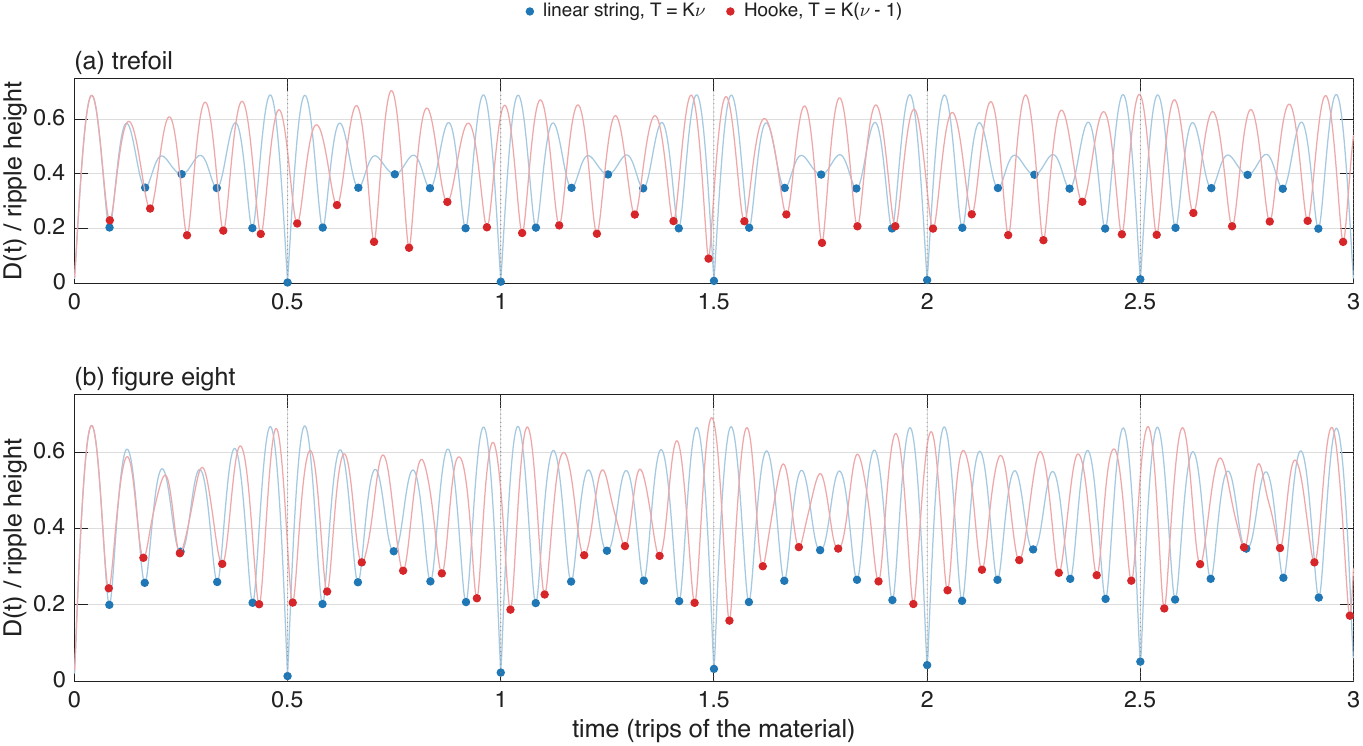}
\caption{Return of a six-wave ripple of height $0.03R$ on a sliding trefoil (a) and figure eight (b) in three trips of the material, 30\% stretch, $N = 960$. Lines, $D(t)$ divided by the ripple height; dots, its minima. Blue, linear string; red, Hooke's law; dotted lines, multiples of half a trip.}
\label{fig:7}
\end{figure*}

Table~\ref{tab:VII} and Fig.~\ref{fig:7} give the results for runs of three trips. With $\lambda = 0$ the trefoil's ripple returns at every half trip. The first return misses by 4.8\% of the ripple height for $N = 240$, 1.2\% for $N = 480$ and 0.3\% for $N = 960$, and with 960 masses the miss grows by 0.3\% at each later return. It falls by a factor of 4 when the spacing is halved, as the chain's $O(h^{2})$ error should, and the continuum return is exact. The figure eight behaves in the same way (4.6\% and 1.4\% at the first return for $N = 480$ and 960). As $\lambda$ increases toward Hooke's law the return time moves away from half a trip and the miss grows, to 5\% at $\lambda = 0.25$ and 9\% at $\lambda = 0.5$ for the trefoil. Under Hooke's law the trefoil's ripple misses by 22\% of its height near half a trip and the figure eight's by 21\% at 30\% stretch, and the trefoil's by 10\% at 100\% stretch, where the subtracted constant is half of $K\nu$ (960 masses). Like the linear ripple, the Hookean ripple comes near its starting shape about every twelfth of a trip, and some later approaches are closer than the one near half a trip. In three trips the closest are 9\% for the trefoil (at 1.49 trips) and 16\% for the figure eight (at 1.54 trips) at 30\% stretch, and 6\% for the trefoil at 100\% stretch (at 2.23 trips). The trefoil values change by less than 1\% of the ripple height between 480 and 960 masses. The figure eight's value is less certain, since the linear figure eight already misses by 3\% at 1.5 trips with 960 masses. A small ripple is nearly a linear motion about the sliding knot, and closer approaches in longer runs cannot be excluded. The return at every half trip, however, is lost under Hooke's law.

\begin{table*}[!htb]
\caption{Return of a six-wave binormal ripple of height $0.03R$ on a sliding knot in runs of three trips of the material. The return near half a trip is the smallest $D(t)$ between 0.45 and 0.55 trips, and the closest approach is the smallest minimum of $D(t)$ after the start. Both are given as fractions of the ripple height, with their times in trips.}
\label{tab:VII}
\begin{ruledtabular}
\begin{tabular}{llllllll}
Knot & Stretch & Tension law & $N$ & Return time & $D$ at return & Closest approach & Its time \\
\hline
Trefoil & 30\% & Linear ($\lambda = 0$) & 240 & 0.5000 & 0.048 & 0.048 & 0.50 \\
Trefoil & 30\% & Linear ($\lambda = 0$) & 480 & 0.5000 & 0.012 & 0.012 & 0.50 \\
Trefoil & 30\% & Linear ($\lambda = 0$) & 960 & 0.5000 & 0.003 & 0.003 & 0.50 \\
Trefoil & 30\% & $\lambda = 0.25$ & 480 & 0.492 & 0.053 & 0.053 & 0.49 \\
Trefoil & 30\% & $\lambda = 0.25$ & 960 & 0.492 & 0.048 & 0.048 & 0.49 \\
Trefoil & 30\% & $\lambda = 0.5$ & 480 & 0.481 & 0.091 & 0.091 & 0.48 \\
Trefoil & 30\% & $\lambda = 0.5$ & 960 & 0.481 & 0.088 & 0.088 & 0.48 \\
Trefoil & 30\% & Hooke ($\lambda = 1$) & 240 & 0.522 & 0.227 & 0.126 & 1.49 \\
Trefoil & 30\% & Hooke ($\lambda = 1$) & 480 & 0.523 & 0.220 & 0.094 & 1.49 \\
Trefoil & 30\% & Hooke ($\lambda = 1$) & 960 & 0.523 & 0.219 & 0.090 & 1.49 \\
Trefoil & 100\% & Linear ($\lambda = 0$) & 480 & 0.5000 & 0.012 & 0.012 & 0.50 \\
Trefoil & 100\% & Linear ($\lambda = 0$) & 960 & 0.5000 & 0.003 & 0.003 & 0.50 \\
Trefoil & 100\% & Hooke ($\lambda = 1$) & 480 & 0.472 & 0.106 & 0.064 & 2.23 \\
Trefoil & 100\% & Hooke ($\lambda = 1$) & 960 & 0.471 & 0.104 & 0.059 & 2.23 \\
Figure eight & 30\% & Linear ($\lambda = 0$) & 480 & 0.5000 & 0.046 & 0.046 & 0.50 \\
Figure eight & 30\% & Linear ($\lambda = 0$) & 960 & 0.5000 & 0.014 & 0.014 & 0.50 \\
Figure eight & 30\% & Hooke ($\lambda = 1$) & 240 & 0.510 & 0.263 & 0.249 & 0.08 \\
Figure eight & 30\% & Hooke ($\lambda = 1$) & 480 & 0.512 & 0.210 & 0.204 & 1.02 \\
Figure eight & 30\% & Hooke ($\lambda = 1$) & 960 & 0.512 & 0.206 & 0.159 & 1.54 \\
\end{tabular}
\end{ruledtabular}
\end{table*}

The difference lies in the wave speeds. For the linear string transverse and longitudinal waves travel at the same speed, and a ripple and the stretching it induces at each bend travel together. For Hooke's law the transverse waves are slower by $[(\nu - 1)/\nu ]^{1/2}$, and the varying curvature and torsion of a knot couple the two. The ring escapes this coupling because its curvature is uniform and its torsion zero, which is why its out-of-plane flappers return in linear theory. Whether exactly periodic knot flappers persist under Hooke's law is open. Hooke's law is a perturbation of an isochronous system, and the surviving orbits could in principle be located by averaging \cite{ref12,ref24}, but the perturbation is not small. A discretization that keeps the relabeling symmetry exact, such as a Fourier series in $s$, would make each sliding knot an exact relative equilibrium and allow its flappers to be sought by Newton's method.

\section{Conclusion}

In summary, I have shown that the spinning ring of a free Hookean string is surrounded by relative periodic orbits that can be computed to residuals below $5 \times 10^{-9}$ and classified by their symmetry and their Floquet multipliers. Out-of-plane rotating waves turn at twice the rotation rate whatever their lobe number. Standing flappers with two to four lobes exist at stretches from 5\% to 100\%, and with five lobes at 30\%. The frequency shift of the saddle flapper changes sign near a stretch of 1/12, where the saddle resonates with the breathing of the ring and two families of flappers overlap. A period-doubled branch carries the saddle flapper continuously into a nearly frozen saddle with a rotating 3-lobe wave, and two period-tripled branches add in-plane ripples. The chain only approximates the string. The orbits and their strong instabilities change little with more masses and are properties of the continuous string. Most weak instabilities, which grow by less than about 25\% per rotation, change with the number of masses and are artifacts of approximating the string by a finite chain. The saddle flapper is neutrally stable at small heights and unstable from $0.25R$ at 30\% stretch, and flappers with more lobes become strongly unstable at smaller heights. Near an unstable saddle flapper the loop exchanges energy back and forth with an in-plane oval. For the linear string every knot carries an exactly periodic flapper; under Hooke's law ripples on sliding knots miss their starting shape by 10\% to 22\% of their height at half a trip, and in three trips come no closer than 6\% to 16\%.

Several questions remain open. Healey's theorems settle the stability of the sliding loops of a stiff string, Hooke's law among them. The spinning ring is orbitally stable \cite{refH90}, and every other sliding loop, the knots among them, is orbitally stable relative to the family of all sliding loops, at any speed \cite{ref29}. His Lyapunov function combines the energy with the circulation, which is the pseudomomentum $J$ up to a constant factor, and his theorem holds for as long as smooth solutions exist. A disturbance can therefore move a sliding knot to a nearby sliding shape but not away from the family, which is consistent with the ripples of Sec.~V, which stay near the knot without returning to their starting shape. The weak instabilities at small heights need a discretization fine enough to separate the modes of an orbit from the short waves of the chain, or a spectral method. Rotating waves of large amplitude need pseudo-arclength continuation and a guard against the nearly sliding family of the chain. The in-plane flapper families, the failed branch switches of Sec. IV~D, and exactly periodic saddle--oval exchange orbits remain to be computed. For knots, a discretization that keeps the relabeling symmetry would allow knot flappers to be sought directly. Contact, gravity and air drag would connect the free loop to the lariat chain \cite{ref5}, the chain fountain \cite{ref30} and the unfurling string \cite{ref31}. Under gravity, Perkins and Mote showed in the laboratory that axial motion can hold up a cable arch that would otherwise collapse \cite{refPM}, and Brun, Ribe and Audoly computed the shapes of a lasso and their stability by numerical continuation \cite{refBrun}.

The free loop is conservative, and its motions near the ring are those of a Hamiltonian system with symmetry. Near a neutrally stable flapper the motion stays close to the orbit, as it would on an invariant torus: the linear start of the saddle holds its peak heights to within 3\% over 40 rotations. Runs started near the strongly unstable 3- and 4-lobe flappers at 30\% stretch leave the orbit within tens of rotations. In the Fermi--Pasta--Ulam lattice, similarly, chaos sets in where simple periodic orbits lose stability \cite{ref33}. Over the next 3000 rotations 10\% to 25\% of the kinetic energy moves to wavenumbers above 20 along the loop, and the largest Lyapunov exponent settles at 4 to 11 per rotation with 120 masses. Doubling the number of masses roughly doubles it, from 6.9 to 14.2 per rotation for the 3-lobe flapper at $0.3R$ and from 11.2 to 20.7 for the 4-lobe flapper at $0.275R$. The chaotic motion of the free loop is therefore set by the shortest waves of the chain. A perfectly flexible string has no shortest wavelength, and its chaotic motion has no finite-dimensional structure that the periodic orbits could describe. A description of chaotic motion in terms of the orbits computed here would need bending stiffness, which supplies a shortest wavelength, and forcing and dissipation, which supply an attractor. A stretched wire driven near its lowest frequency, with damping, shows what forcing and dissipation do: it whirls, and its route to chaos in the geometrically exact equations is reproduced by a model with one transverse mode in each plane \cite{ref34}. I believe the free loop, with its exact sliding family and its simple equations, makes an excellent laboratory for the periodic orbits of a conservative continuum with symmetry.

\section*{Data availability}

The MATLAB code and the data behind the figures and tables are openly available from Zenodo \cite{refZenodo}. The movies are provided as Supplemental Material \cite{refSM}.

\begin{acknowledgments}
The author used Claude (Anthropic) to document and test the simulation code, check algebra, edit the text for clarity, and assist with figures. The author reviewed and validated all AI-assisted output and takes full responsibility for the content.
\end{acknowledgments}

\end{document}